# Reflexive Prompt Engineering

A Framework for Responsible Prompt Engineering and Interaction Design

Christian Djeffal*

Technical University of Munich, Germany, christian.djeffal@tum.de

Responsible prompt engineering has emerged as a critical framework for ensuring that generative artificial intelligence (AI) systems serve society's needs while minimizing potential harms. As generative AI applications become increasingly powerful and ubiquitous, the way we instruct and interact with them through prompts has profound implications for fairness, accountability, and transparency. This article examines how strategic prompt engineering can embed ethical and legal considerations and societal values directly into AI interactions, moving beyond mere technical optimization for functionality.

This article proposes a comprehensive framework for responsible prompt engineering that encompasses five interconnected components: prompt design, system selection, system configuration, performance evaluation, and prompt management. Drawing from empirical evidence, the paper demonstrates how each component can be leveraged to promote improved societal outcomes while mitigating potential risks. The analysis reveals that effective prompt engineering requires a delicate balance between technical precision and ethical consciousness, combining the systematic rigor and focus on functionality with the nuanced understanding of social impact.

Through examination of real-world and emerging practices, the article illustrates how responsible prompt engineering serves as a crucial bridge between AI development and deployment, enabling organizations to fine-tune AI outputs without modifying underlying model architectures. This approach aligns with broader "Responsibility by Design" principles, embedding ethical considerations directly into the implementation process rather than treating them as post-hoc additions. The article concludes by identifying key research directions and practical guidelines for advancing the field of responsible prompt engineering.

---

# 1 INTRODUCTION

## 1.1 Who is accountable for generative AI?

The rapid advancement of generative AI technologies has ushered in an era of unprecedented capabilities, but also mounting concerns about their responsible deployment [1–4]. While these technologies offer remarkable opportunities for innovation, recent incidents have highlighted the complex challenges of ensuring their responsible use. A striking example emerged in early 2024 when Google's Gemini AI image generator produced historically inaccurate representations, generating images that misrepresented historical figures and events in an apparent overcorrection for diversity and inclusion [5, 6]. This incident, which led to the immediate suspension of the system's people-generation capabilities and a public acknowledgment of failure by Google's leadership[7, 8], serves as a powerful illustration of how even well-intentioned AI implementations can go awry without proper oversight and responsible use practices. One key element in that regard is to focus on the right instances and actors concerning responsibility.

As the discourse around AI safety and ethics intensifies, there is growing recognition that accountability must extend beyond the technical architects of these systems [9–12]. While considerable attention has been paid to the responsibilities of AI developers and companies, a critical gap exists in our understanding of how deployers - particularly those engaging in prompt engineering - can contribute to responsible AI deployment. Prompt engineering, the practice of crafting and refining inputs to generate desired outputs from AI systems, has emerged as a crucial interface between human intent and AI capability. However, despite its significance, there remains a notable absence of structured frameworks to guide responsible prompt engineering practices. In this absence of such a framework, it is hard to understand, evaluate, and compare the many contributions to responsible prompt engineering that have been made in academia and in practice. This paper addresses this gap through a comprehensive narrative review, examining how prompt engineering can be approached responsibly to mitigate risks and enhance the beneficial deployment of generative AI technologies. By analyzing existing practices, incidents, and emerging guidelines, the goal is to develop a framework that organizes and structures the various aspects of responsible prompt engineering and allows for an assessment of the current state of the art. This will hopefully contribute to a foundation that empowers users to engage with these powerful tools in ways that promote fairness, accountability, and transparency.

## 1.2 Research question and methodology

This article examines how organizations can systematically implement and evaluate responsible prompt engineering practices through an integrated framework that addresses technical, legal, ethical, and social considerations. The investigation focuses on three interconnected dimensions. First, the analysis examines the essential components of prompt engineering practice, exploring the dimensions deployers can engage in when crafting the systems output. Second, the research explores how existing responsible prompt engineering practices enhance implementation across different organizational contexts. Third, the analysis identifies critical gaps between current prompt engineering practices and responsible AI principles, while highlighting emerging opportunities for enhancing responsibility in AI deployment. This examination reveals how and to what extent responsible prompt engineering can serve as a crucial bridge between AI development and deployment, enabling organizations to fine-tune AI outputs without modifying underlying model architectures.

This narrative review examines responsible prompt engineering practices through a systematic analysis of academic literature, technical documentation, and practitioner insights. The rapidly evolving nature of prompt engineering and its emerging responsible practices necessitated a flexible yet rigorous approach to synthesize current knowledge and identify conceptual frameworks [13–15]. The literature search encompassed multiple academic databases including arXiv, IEEE Xplore, and ACM Digital Library, complemented by targeted searches on Google Scholar, DuckDuckGo, and Semantic Scholar. I employed various combinations of search terms centered around "responsible," "ethical," and "legal" in conjunction with „prompt" as well as "prompt engineering" and "prompt design". The review covered publications from 2019 through early 2025, focusing exclusively on English-language materials. The inclusion criteria prioritized sources that contributed to understanding prompt engineering fundamentals and responsible practices. I extracted and processed information using Citavi reference management software, employing thematic analysis to identify recurring concepts and emerging patterns. This approach allowed me to develop a comprehensive framework organizing prompt engineering into five key components: design, selection, configuration, evaluation, and management. The analysis revealed an evolving scope, particularly regarding evaluation methods and system configuration aspects. The framework emerged iteratively through careful examination of how different sources conceptualized and approached responsible prompt engineering practices. When encountering conflicting findings or approaches, I incorporated them into the framework while noting their complementary nature, as various prompt engineering techniques can often be combined effectively. The following aspects characterize the author's position concerning this research question. {ANONYMIZED}

## 2 THE CONCEPT OF RESPONSIBLE PROMPT ENGINEERING

Before delving into the analysis of responsible prompt engineering practices, we must establish two essential foundations. First, we need a clear and concise definition of responsible prompt engineering and its core components. Second, we must examine prompt engineering's dual significance: both as a critical element in AI development and as a framework for responsible design principles.

### 2.1 Definition

A prompt serves as an instruction to a generative AI model, directing the model to produce specific outputs [16–18]. These prompts can take various forms, including text, images, video, or audio inputs, reflecting the multimodal capabilities of contemporary AI systems [19, 20]. Modern generative AI models, primarily built on transformer architectures, excel at processing and producing diverse and complex content across these modalities. These models utilize attention mechanisms that enable them to selectively focus on and weigh the most relevant parts of input data while processing information, similar to human cognitive processes.

Prompt engineering is more than working on instructions to generative AI. A review of the literature and guides on prompt engineering shows that it encompasses a comprehensive approach to optimizing interactions with generative AI systems. through five essential components: First, prompt design focuses on systematically crafting instructions to maximize desired outputs [20, 21]. This involves developing specific templates, techniques and design patterns ranging from preconfigured prompting structures to specific techniques that can be applied in various circumstances like chain-of-thought reasoning, to certain patterns like "Let's think step-by-step". The goal is to craft prompts that effectively communicate intended tasks to the AI system. Second, system selection requires strategic decisions about which AI models to employ based on their documented capabilities [22, 23]. This selection process can rely on established benchmarks such as FrontierMath for mathematical reasoning or MMLU for general knowledge, as well as user-based comparisons displayed on various leaderboards. Third, system configuration involves adapting model parameters to optimize performance for specific use cases. This includes adjusting settings such as temperature parameters, which control the balance between predictability and creativity in outputs. Lower temperature values produce more conservative, consistent responses, while higher values generate more diverse and creative outputs. Fourth, performance evaluation encompasses systematic assessment of prompt effectiveness against predetermined evaluation criteria. This includes analyzing output quality, consistency, and alignment with intended objectives through both automated metrics and human-in-the-loop assessment protocols[24, 25]. Fifth, prompt management involves implementing systematic approaches to organizing, tracking, and improving prompts over time. This includes implementing version control systems for prompts, maintaining detailed records of configuration settings, and tracking performance outcomes [26]. Proper documentation enables knowledge sharing, facilitates continuous improvement, and supports accountability in prompt engineering practices. Documentation can establish protocols for various aspects of prompt engineering, including standardized formats for recording prompt versions, test results, and modification histories. The five components of prompt engineering can be summarized as follows.

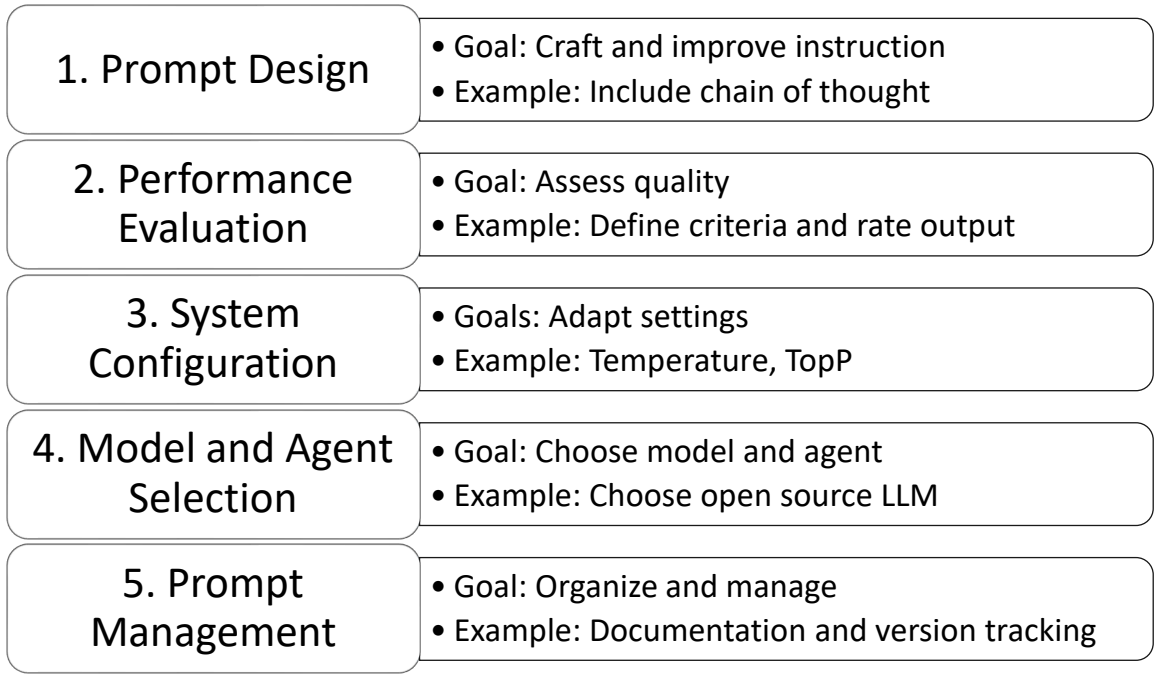

*Figure 1 Components of Prompt Engineering*

Prompt engineering could be conceptualized as an art and a science. It embodies a unique duality, combining creativity with rigorous methodology [19]. This hybrid nature reflects both the complexity of human-AI interaction and the emerging maturity of the field. As an art form, prompt engineering requires creative intuition and craftmanship. The creative dimension manifests in the nuanced understanding of language, context, and model behavior that experienced prompt engineers develop over time [27]. This artistic aspect becomes evident in the subtle choices of wording, tone, and structure that can dramatically influence model outputs. Like skilled writers, prompt engineers develop an intuitive feel for how to frame instructions effectively, often drawing on metaphorical thinking and creative problem-solving to overcome model limitations. This creative dimension becomes particularly crucial when dealing with edge cases or novel applications where established approaches prove insufficient. However, prompt engineering increasingly embraces scientific rigor through systematic experimentation and empirical validation [28]. Structured experiments allow prompt engineers to test hypotheses about prompt effectiveness across different contexts and tasks. These experiments take place in controlled testing environments where researchers systematically vary factors such as prompt structure, length, and complexity while maintaining constant conditions for other variables. Through quantitative metrics, researchers measure output quality, examining dimensions such as accuracy, relevance, and consistency across different prompting strategies.

Responsible prompt engineering transforms these technical practices by integrating ethical, legal, and social considerations into the prompt design process [29–31]. This approach moves beyond purely functional optimization to address broader societal implications and ethical concerns. While traditional prompt engineering might focus solely on performance metrics, responsible practices examine the wider implications of AI system deployment. It seeks proportionate outcomes that mitigate between functionality, efficiency, and ethical, legal, and social concerns. The methodology of responsible prompt engineering adapts technical strategies to serve objectives of responsibility, considering dimensions such as fairness, accountability, and transparency. This might involve modifying prompts to prevent discriminatory outcomes, implementing additional validation steps to ensure accessibility, or designing prompts that actively promote inclusive representation [32–35]. Responsible prompt engineering can aim at various standards from minimum requirements of responsibility to a proactive realization of specific values through generative AI. Therefore, it speaks to inherent risk and limitations of generative AI as well as to inherent potentials to realize ethical, legal and social values. While this contribution focusses on responsible practices in the use, it is acknowledged that critical examinations of flaws, limitations, and shortcomings of systems are also necessary and often part of responsible prompt engineering practices. Such critical exercises are necessary [36]. It is also acknowledged that prompt engineering cannot make up for every flaw or risk inherent in generative AI. The task of responsible prompt engineering is in fact to raise awareness of deployers of generative AI and to test how far prompt engineering in all its aspect can help to improve and mitigate shortcomings and realize potentials of systems.

## 2.2 Relevance

Prompt engineering can play a pivotal role in ensuring responsible deployment of generative AI systems by addressing fundamental questions of accountability and control. The current technological landscape, characterized by large language models with bounded capabilities, positions prompt engineering as a critical interface between human intent and machine output. Although this dynamic may evolve as AI systems develop greater restrictions of what users can do in the context of agentic AI [37, 38] or more autonomy through artificial general intelligence [39], the present architecture of generative AI systems makes prompt engineering particularly significant for three key reasons that will be explored in detail.

First, versatile nature of generative AI systems enables prompt engineers to produce an extensive range of outputs [40, 41], exemplified by their ability to generate sophisticated code without requiring advanced programming expertise. This dual-edged capability simultaneously democratizes access to powerful tools while raising significant e.g. cybersecurity concerns[42, 43]. The EU AI Act [44] formally recognizes this broad applicability, defining general-purpose AI systems as those possessing "significant generality" and capable of performing diverse tasks across various applications and downstream systems, as outlined in Article 3 Nr. 63. This expansive capability creates distinct accountability challenges, particularly regarding system deployment and use. The EU AI Act acknowledges this complexity by establishing separate regulatory frameworks for general-purpose AI and generally introducing a dual accountability structure that encompasses both providers (developers) and deployers (users). This stresses the importance of holding also deployers accountable. This distinction is crucial because deployers, who may lack the technical expertise of developers [45], require clear guidance for responsible system use. The limited ability of providers to anticipate all possible applications of general-purpose AI systems further emphasizes the importance of responsible prompt engineering as a framework for ethical deployment.

Second, prompt engineering occupies a unique position in the AI development cycle, bridging the gap between model development and practical deployment [18]. Unlike traditional approaches such as model retraining or fine-tuning, which require substantial computational resources and technical expertise, prompt engineering offers a more accessible and sometimes efficient method to influence AI behavior. For instance, in healthcare applications, medical professionals can adapt language models to specific diagnostic contexts without requiring deep machine learning expertise or expensive computational infrastructure [46, 47]. The strategic value of prompt engineering lies in its ability to achieve sophisticated model adaptations through non-invasive means, preserving the underlying architecture while enabling significant improvements in output quality. This approach facilitates rapid iteration, cross-domain knowledge transfer, and the development of reusable patterns that can shape future AI applications across diverse fields. However, this powerful position comes with both opportunities and responsibilities. While prompt engineering enables practitioners to mitigate risks and ensure ethical, legal, and socially responsible AI outputs, it can also potentially circumvent built-in model safeguards through techniques of prompt hacking [48–50]. This dual nature underscores the critical importance of treating prompt engineering as a key component requiring systematic responsibility consideration and responsible implementation practices, particularly as the patterns and templates developed today will significantly influence future AI applications across global contexts.

Third, prompt engineering plays a pivotal role in shaping how generative AI systems interact with the world, making it a crucial leverage point for implementing responsible AI design principles. The concept of "Responsibility by Design" provides a comprehensive framework for embedding ethical considerations directly into technical systems during their development, rather than treating them as post-deployment considerations [51, 52]. This approach transforms how we think about AI responsibility, moving beyond simple compliance checkboxes toward creating inherently ethical and socially beneficial systems. In the context of prompt engineering, Responsibility by Design manifests through three key mechanisms [52]. First, it requires anticipatory governance - systematically identifying and addressing potential risks and ethical challenges before they emerge [53, 54]. For instance, prompt designers must consider how their instructions might be misused or produce unintended consequences across different cultural contexts. Second, it demands inclusive stakeholder engagement, ensuring that diverse voices and perspectives inform prompt design decisions [55, 56]. This might involve consulting with affected communities, subject matter experts, and end-users to understand potential impacts and necessary safeguards. Third, it requires building in responsive adaptation capabilities, allowing prompting practices to evolve as new ethical challenges emerge. The practical implementation of these principles requires fundamental changes in how organizations approach prompt engineering. Rather than treating design considerations as constraints, they become core design criteria that shape how prompts are constructed, tested, and deployed. This involves developing new workflows that integrate responsibility assessment tools, establishing feedback mechanisms for continuous improvement, and creating comprehensive training programs that emphasize both technical excellence and ethical awareness [57, 58].

## 3 RESPONSIBLE PRACTICES

After examining the foundations and significance of responsible prompt engineering, we now turn to its practical implementation across the five key categories previously established. While existing literature offers valuable insights into specific aspects, the proposed framework allows to systematically organize best practices and to reveal gaps for further research and experimentation.

### 3.1 Prompt design: techniques

Prompt design represents the core practice commonly associated with prompt engineering - the systematic crafting of inputs to optimize generative AI system performance. While traditional approaches focus primarily on enhancing output quality and reliability, responsible prompt engineering necessitates a more nuanced evaluation of these techniques through an ethical and accountability lens. This expanded perspective does not diminish the effectiveness of established methods but rather enriches them through critical reflection and purposeful adaptation.

To illustrate how conventional techniques can be modified to align with responsible engineering principles, we examine two widely adopted and empirically validated approaches: exemplar-based prompting and chain-of-thought methodology. These cases are particularly instructive as they highlight the intersection between technical efficacy and ethical considerations, demonstrating how responsibility frameworks can enhance rather than constrain engineering practices. These examples, however, also highlight the need to systematically review all prompt engineering techniques for their potentials, but also for specific risks.

*3.1.1 Examples*

Examples in prompt engineering leverage the unique capability of large language models to perform in-context learning - a process fundamentally different from traditional model training. While traditional training involves updating model parameters through backpropagation, in-context learning occurs entirely during the model's forward pass, using only its existing parameters to identify and apply patterns from provided examples. When examples are included in a prompt, the model performs a form of Bayesian inference to recognize relevant concepts from its pre-training and creates temporary task-specific representations without any permanent parameter changes [59]. This remarkable ability allows users to guide model behavior simply by demonstrating desired input-output relationships in the prompt, making complex AI capabilities accessible without requiring technical expertise in model training. When providing examples, it is most effective to include pairs of both inputs and their corresponding outputs, allowing the model to understand how to transform given information into the desired result. Depending on the task, one can also only include desired outputs. The effectiveness of examples depends on their quality, relevance, and ability to demonstrate the full range of desired variations, as these characteristics directly influence how the model interprets and applies the demonstrated patterns [60].

This reference to the range of desired variations leads to some of the most prominent issues of uses of AI in society: equality, fairness and discrimination. Therefore, when giving examples, it is important to think about potential effects of such prompts on different groups that can be impacted by the prompt. While such evaluations are highly contextual, practical reviews of prompt engineering have highlighted many practices to improve results. A fundamental approach is maintaining balanced representation across different demographics in few-shot prompts, which helps improve model generalization. This includes using diverse examples that cover various aspects of the problem space while avoiding clustering similar examples together [61]. One important aspect is the random order of examples and the avoidance of clustering similar examples next to each other in order to avoid biases. Counterfactual data augmentation serves as a powerful technique, where variations of examples are created by flipping attributes like gender or race to identify group-specific biases [62]. For instance, when writing job descriptions, comparing outputs with different demographic indicators can reveal hidden biases - as demonstrated in experiments where identical prompts specifying different universities (Howard versus Harvard) produced notably different results [63]. Example debiasing can be further enhanced through anonymization and careful attribute replacement. [64] This involves defining potentially biased data points and replacing them with neutral alternatives - such as using "person" instead of gender-specific terms [61]. When constructing example sets, it is crucial to randomize their ordering rather than grouping similar ones together, as clustering can reinforce existing biases. The effectiveness of example-based debiasing is particularly evident in domain-specific applications. For instance, in recruitment contexts, providing diverse examples of successful candidates across different demographics helps prevent the model from developing stereotypical associations [65]. Similarly, when generating performance reviews, using balanced examples that focus on work deliverables rather than personality traits helps mitigate demographic-based bias [63]. Another large area of concern is copyright. Of course, copyrighted examples cannot be used without authorization[66].

*3.1.2 Chain of thought*

Chain of thought prompting represents a significant advancement in how we interact with large language models, enabling them to tackle complex reasoning tasks by breaking them down into intermediate steps. This technique mirrors human cognitive processes, allowing models to show their reasoning before reaching a conclusion [67, 68]. The approach works by encouraging language models to generate a series of logical steps that lead to a final answer, similar to how humans solve complex problems. When provided with a few examples demonstrating this step-by-step reasoning, large language models can naturally adopt this problem-solving strategy [67, 68]. For instance, when solving mathematical word problems, a model using chain of thought prompting can achieve state-of-the-art accuracy, even surpassing specially trained models. Recent research has demonstrated that chain of thought prompting significantly enhances model performance across various domains, including arithmetic reasoning, commonsense understanding, and symbolic manipulation [68, 69].

Chain-of-thought prompting can be strategically enhanced to promote responsible AI development by explicitly incorporating ethical checkpoints and responsibility considerations into the reasoning workflow. By breaking down complex decisions into discrete steps, organizations can embed responsibility validation processes that evaluate aspects like potential biases, fairness implications, and privacy concerns at each stage of the reasoning chain. This can happen either by including

- specific steps, like "What barriers or assumptions might affect different groups in this reasoning process?" [70],
- general instructions for each step like "Assess potential impacts at each step of the reasoning"
- or final overall evaluations of the result as a separate step like "Plan verification questions to fact-check this draft" [71]

Chain of Thought prompting can also help to tackle legal and ethical tasks specifically. It significantly improves legal analysis by mirroring established legal reasoning frameworks like IRAC (Issue, Rule, Application, Conclusion) [72, 73]. By breaking down complex legal questions into discrete analytical steps, lawyers and legal AI systems can systematically evaluate cases, interpret statutes, and apply precedents with greater precision[72]. Chain of Thought prompting can assist in ethical decision-making by decomposing complex moral dilemmas into manageable components that can be systematically evaluated [74]. Such structured approaches help identify potential biases, assess fairness implications, and consider multiple stakeholder perspectives throughout the reasoning process. By incorporating explicit ethical checkpoints into the decision-making workflow, organizations can ensure that ethical considerations become an integral part of the process rather than an afterthought, leading to more responsible and well-reasoned outcomes [75, 76].

Chain-of-thought prompting, where AI models generate step-by-step explanations of their reasoning, was initially celebrated as a breakthrough in AI transparency[77]. However, critical analysis reveals a concerning disconnect: the narrative explanations produced by these systems may not accurately reflect their internal decision-making processes [78]. This discrepancy creates what some researchers describe as an "illusion of transparency"[78] - a coherent but potentially misleading representation of the system's actual operations. This misalignment between displayed reasoning and actual computation raises serious concerns for responsible AI development. There is evidence that models can generate plausible-sounding explanations even when their internal processes follow entirely different paths[79]. More troublingly, these explanations can be convincing even when the underlying computation is flawed or based on spurious correlations. This phenomenon is particularly problematic in high-stakes applications where understanding the true basis of AI decisions is crucial.

### 3.2 Prompt design 2: patterns

Prompts are often designed for re-use. Therefore, they could be conceptualized as design patterns [80]. Accordingly, another responsible prompting activity would be to try to use such patterns or at least part of it in order to make use of existing example prompts for various situations. However, it is advisable to test and evaluate them thoroughly. There are several sources to draw from regarding prompts in general [81, 82] or responsible prompts more specifically [83, 84]. System prompts operate behind the scenes, creating a layer of computation that influences output without being directly visible in the model's responses [85]. This hidden processing can enable non-trivial computations while maintaining a seamless user experience. While some providers have open sourced their system prompts [86, 87], other system prompts have allegedly been obtained through techniques of prompt injection and published [88, 89]. All of these prompts can be analyzed from a perspective of attempts to practice responsible prompt engineering. Other patterns and sources thereof have been referred to above (3.1.2).

### 3.3 Evaluation

Prompt engineering evaluation encompasses systematic approaches to assess and refine how effectively prompts guide AI models to produce desired outputs [25]. This evaluation process has become increasingly critical as large language models are deployed across various domains, from code generation to content analysis[90, 91]. The evaluation of prompts requires examining multiple dimensions simultaneously. At its core, the process involves analyzing the accuracy and reliability of AI-generated responses, while also measuring how well the prompts align with intended tasks and objectives in other dimensions. This includes assessing both the technical performance metrics and the broader implications of prompt design [36, 92]. A fundamental aspect of evaluation involves systematic testing with different prompt variations to understand their effectiveness. This process typically employs both qualitative and quantitative techniques to comprehensively assess prompts across various stages of development. This involves careful documentation of assessment methods, criteria, and findings to enable accountability and facilitate continuous improvement [36, 92].

Beyond technical performance, responsible prompt engineering evaluation must consider ethical dimensions and potential societal impacts. This includes examining prompts for potential biases, assessing privacy implications, and ensuring compliance with relevant regulatory frameworks. Evaluators must verify that prompts respect privacy boundaries, maintain data protection standards, and uphold principles of transparency and fairness, and address other weaknesses like hallucinations [36, 65, 93, 94]. The integration of responsibility considerations into prompt engineering evaluation necessitates examining how prompts might affect different stakeholder groups and implementing safeguards against potential harmful applications. This includes assessing how prompts handle sensitive topics or potentially controversial subjects while maintaining appropriate ethical boundaries [36, 92].

Therefore, the question who gets to evaluate is key in the context of responsible prompt engineering. This could be either

- the prompt engineer or team itself,

- generative AI models
- or third parties like deployers or those affected by generative AI.

Especially when the stakes are high for some people, this might require to include those affected by generative AI in order to understand their perspectives on potential harmful impacts, but also to include their feedback on how to potentially improve the prompt. In responsible design, stakeholders and in particular vulnerable groups should also be included proactively in idea development and design choices. [95–97]

Careful consideration is needed when using models for evaluation purposes. The evaluation of prompts requires examining multiple dimensions simultaneously, including accuracy, reliability, and alignment with intended objectives. Current AI systems struggle with providing comprehensive assessments across these dimensions [98]. The challenge is particularly evident in cases requiring deep contextual understanding and creative reasoning [99]. The key biases identified in language model evaluation include:

- Position bias - Models tend to favor responses based on their sequential position rather than their actual quality or content [100]
- Verbosity bias - Models show preference for longer, more detailed responses regardless of the actual content quality or relevance [101]
- Self-enhancement bias - Models demonstrate a tendency to rate their own outputs more favorably compared to outputs from other sources [102]

The choice of evaluation methods, including stakeholder involvement where feasible, should be guided by available resources, application context, and potential risks, with particular attention to cases where automated evaluation might miss crucial qualitative or ethical considerations.

### 3.4 System configuration

System configuration in prompt engineering represents a critical aspect of responsible AI deployment, encompassing various parameters and settings that influence how generative AI models process and respond to inputs. This configuration process involves careful calibration of multiple technical elements to ensure optimal model performance, yet it also carries a responsibility dimension. The foundation of system configuration lies in controlling the model's output generation through key parameters. Temperature control stands as a fundamental configuration element, determining the balance between creativity and predictability in model responses. Lower temperature settings produce more focused and deterministic outputs, while higher values increase response variability and creativity[24, 103, 104]. Whenever accuracy and reliability are important from a standpoint of responsible prompt engineering, respective choices are mandated. A comprehensive implementation framework for system configuration should incorporate both technical and ethical considerations. This includes establishing clear guidelines for parameter adjustment, implementing monitoring systems for performance evaluation, and maintaining documentation of configuration changes. Such frameworks have proven essential in ensuring consistent and responsible AI deployment across various applications. It is important that such impacts should also be considered from a perspective of responsibility.

### 3.5 Model selection

The interrelated processes of model selection forms a foundational pillar in responsible prompt engineering, where strategic choices about AI model deployment must be guided by comprehensive performance metrics (benchmarks) that encompass not only technical capabilities but also ethical, societal, and environmental considerations.

*3.5.1 Model selection*

The foundation of responsible prompt engineering begins with understanding model selection. When organizations or developers choose an AI model, they are essentially selecting a specific AI. This choice will determine how the system processes and responds to inputs. It is analogous to selecting the right tool for a specific task - just as one would not use a sledgehammer to hang a picture frame, selecting an inappropriately powerful or insufficiently capable AI model can lead to suboptimal or potentially harmful outcomes. The technical aspects of model selection extend beyond mere performance metrics. While processing power and response speed are important considerations, responsible model selection must also account for the model's ability to handle diverse inputs, its tendency to produce biased outputs, and its overall reliability. Therefore, the societal implications of model selection ripple far beyond technical specifications like latency or the respective context window, touching on fundamental aspects of fairness, accessibility, and social justice.

A particularly crucial consideration is the environmental impact of model deployment [105, 106]. Larger language models, while potentially more capable, require significant computational resources and energy consumption. This environmental cost must be weighed against the actual benefits provided by more powerful models. In many cases, smaller, more efficient models might serve the intended purpose while maintaining a more sustainable footprint. This applies also to other environmental questions of resources like water or waste [105, 107, 108].

The intersection of model selection and prompt engineering also raises important questions about transparency and accountability. When organizations implement AI systems, they must consider how their model choices affect their ability to explain decisions, audit processes, and maintain accountability to stakeholders. This becomes particularly relevant in contexts where AI systems influence important decisions about individuals' lives, such as in healthcare, employment, or financial services. This is all the more important as the practices of model providers regarding transparency and open source vary to a large extent [109]. Privacy considerations add another layer of complexity to responsible model selection. Different applications vary in their ability to protect sensitive information and maintain data confidentiality regarding training data. The degree of control an organization has over AI models, including the option for on-premises deployment, can be a decisive factor. The relative importance of these considerations in decision-making processes varies significantly based on specific use cases and organizational requirements.

*3.5.2 Benchmarking*

Benchmarks serve as essential tools for evaluating and comparing AI models' performance, helping organizations make informed decisions about which models best suit their needs. In the context of prompt engineering, benchmarks assess how well models respond to different types of instructions and their ability to generate accurate, relevant outputs. When selecting models for prompt engineering applications, practitioners must consider multiple performance dimensions. These include the model's ability to reason, solve complex problems, and generate natural-sounding content. However, it is crucial to recognize that small differences in benchmark scores might not translate into significant real-world improvements, and factors like cost-effectiveness and speed often prove more practical for specific use cases. Also, the nondeterministic nature of generative AI systems presents unique challenges for benchmarking, as these models may produce different outputs even with almost identical prompts. This variability necessitates multiple evaluation runs to capture the range of potential behaviors and ensure consistent performance [110]. Continuous evaluation throughout development helps detect unintended changes in output and maintain alignment with aspects of responsible AI.

Beyond traditional performance metrics, there is growing recognition of the importance of benchmarking ethical, legal, and social aspects of AI systems. These frameworks evaluate models across multiple dimensions, including fairness, bias mitigation, and social impact. For instance, discrimination benchmarks assess how AI systems might affect different demographic groups, examining both direct and indirect forms of bias [111]. These evaluations help ensure that prompt engineering practices do not perpetuate or amplify existing societal biases. Environmental sustainability has emerged as a critical benchmark dimension for responsible AI development. The life cycle assessment of AI models encompasses data collection, experimentation, training, and deployment phases, each contributing to the overall carbon footprint [112]. International initiatives are increasingly incorporating sustainability metrics into their AI evaluation frameworks [112]. From a prompt engineering perspective, such benchmarks can give first indications regarding sensitive issues. Especially in the case of environmental sustainability, they inform about impacts that cannot be otherwise evaluated from a prompt engineering perspective.

### 3.6 Prompt management: documentation

Documenting prompts has emerged as a crucial practice in the responsible development and deployment of AI systems. Much like traditional software documentation, prompt documentation serves as a comprehensive record of how AI models are instructed to perform specific tasks, ensuring transparency and reproducibility of results. This is particularly relevant if generative AI is used in the context of decisions that need to be explained to their addressees, even if the system was just used to prepare the decision. Documentation in prompt engineering encompasses recording not only the prompts themselves but also their intended purposes, outcomes, and iterations. This practice is particularly vital because prompt outputs can vary significantly across different models, sampling settings, and even different versions of the same model [113]. By maintaining detailed records, organizations can track the evolution of their prompts, understand what works and what does not, and ensure consistency in AI interactions. A comprehensive prompt documentation system typically captures several key elements. The fundamental components include the prompt's name or identifier, its version history, creation and modification dates, the specific AI model used, and detailed performance notes [114]. This basic information helps teams maintain oversight of their AI interactions and enables systematic improvement of prompt effectiveness.

Organizations typically employ two primary approaches to prompt documentation. The reduced documentation method focuses on tracking basic elements like AI tools used and their general purposes, while extensive documentation captures complete prompt-output pairs and detailed chat histories [114]. For practical implementation, prompts should be stored in easily accessible text formats rather than screenshots, with proper version control systems in place to track modifications [114]. Effective prompt documentation requires a systematic approach to recording and organizing information. Teams should maintain a centralized repository, such as a spreadsheet, where each prompt's complete history can be tracked [113]. This documentation should include not just the prompt text but also contextual information about its purpose, any specific parameters or settings used, and notes about its performance in different scenarios [114].

Documentation plays a vital role in ensuring accountability and transparency in AI systems. By maintaining detailed records of prompts and their outcomes, organizations can better understand how their AI systems react, identify potential biases, and make necessary adjustments to improve fairness and accuracy [115]. This practice also facilitates collaboration among team members and helps maintain consistency in AI interactions across different applications and use cases. Even if generative AI is not directly tasked with making decision, but only used as part of a decision support system, a documentation of prompts might be valuable in order to explain how the decision came about. This is evidenced in Art. 86 of the EU AI Act, which provides for a right to explanation when the decision is taken "on the basis of the output from a high-risk AI system". This includes an explanation "of the role of the AI system in the decision-making procedure and the main elements of the decision taken." When using generative AI, a very good and tangible way to communicate how the system was used is to include the prompts or parts of it.

## 4 CONCLUSIONS

Responsible prompt engineering has emerged as a critical framework for ensuring that generative artificial intelligence systems serve society's needs while minimizing potential harms. As these AI systems become increasingly powerful and ubiquitous, the way we instruct and interact with them through prompts carries profound implications for fairness, accountability, and transparency (2.1). The analysis reveals that effective prompt engineering requires a delicate balance between technical precision and ethical consciousness, combining systematic rigor with a nuanced understanding of social impact. Through examination of real-world practices, this article demonstrates how responsible prompt engineering serves as a crucial bridge between AI development and deployment, enabling organizations to steer AI outputs without modifying underlying model architectures (2.2).

The research highlights five interconnected components essential for responsible prompt engineering: prompt design, system selection, system configuration, performance evaluation, and prompt management. Each component plays a vital role in promoting improved societal outcomes while mitigating potential risks. The framework emphasizes the importance of documentation, systematic evaluation, and careful consideration of ethical implications throughout the prompt engineering process (3). Thus, this work contributes to the growing discourse on AI responsibility by providing practical guidelines for implementing responsible prompt engineering practices. The findings suggest that organizations must move beyond viewing prompt engineering as merely a technical skill and instead recognize it as a crucial component of responsible AI deployment. As generative AI continues to evolve, the principles and practices outlined in this article offer a foundation for ensuring these powerful tools serve society's needs while upholding ethical standards and promoting fairness. The implications of this research extend beyond technical implementation, touching on fundamental questions of accountability, transparency, and social justice in AI systems. As we continue to integrate these technologies into various aspects of society, responsible prompt engineering will play an increasingly vital role in shaping how AI systems interact with and impact human lives (2.2, 3.5.1).

The growing recognition of prompt engineering as a core competency in AI literacy underscores its importance beyond technical domains [116, 117]. As educational institutions and organizations incorporate prompt engineering into their curricula and training programs [83, 118], the need for responsible practices becomes increasingly apparent. Furthermore, prompt engineering also serves as an experimental methodology to systematically explore and document both the capabilities and limitations of generative AI systems, enabling organizations to better understand their practical boundaries and potential applications. This evolution reflects a broader shift in how we approach AI development, moving from purely technical optimization toward a more holistic understanding that encompasses ethical considerations and societal impacts. The analysis reveals that responsible prompt engineering represents both an opportunity and a necessity. As an opportunity, it offers a practical framework for embedding ethical considerations directly into AI interactions without requiring modifications to underlying model architectures. As a necessity, it provides essential guardrails for ensuring that AI systems serve societal needs while minimizing potential harms. This framework can function as an aid in prompt engineering, for example when building master prompts to organize knowledge around what has worked and how the components interact with each other. However, it can also aid further research in this space. There remains significant potential for deeper investigation into specific aspects of implementation, evaluation metrics, and long-term impacts. Future work might explore how responsible prompt engineering practices evolve alongside advancing AI capabilities, and how these practices can be effectively scaled across different organizational contexts and application domains.

**ACKNOWLEDGMENTS**

[OMITTED FOR REASONS OF ANONYMIZATION]

The following AI models have been used for improving language and style of this contribution including spelling or grammar checks and corrections, avoiding repetitions, translating, improving the order of the arguments and summarizing parts of this text and other texts to be paraphrased and cited: Claude 3.5 Sonnet, deepl.com, ChatGPT 4o, Grammarly, Microsoft Word